\documentclass[aps,prl,twocolumn,superscriptaddress]{revtex4-1}

\usepackage{amsmath,braket,amssymb,natbib,graphicx,float,amsthm,xcolor,outlines,mathrsfs,multirow,hhline}
\usepackage{hyperref}
\usepackage{url}

\begin{document}

\title{Entanglement growth via splitting of a few thermal quanta}
\author{Pradip Laha}
\email{plaha@uni-mainz.de}
\affiliation{Department of Optics, Palack\'{y} University, 17. listopadu 1192/12, 771 46 Olomouc, Czech Republic}
\author{Darren W. Moore}
\email{darren.moore@upol.cz}
\affiliation{Department of Optics, Palack\'{y} University, 17. listopadu 1192/12, 771 46 Olomouc, Czech Republic}
\author{Radim Filip}
\email{filip@optics.upol.cz}
\affiliation{Department of Optics, Palack\'{y} University, 17. listopadu 1192/12, 771 46 Olomouc, Czech Republic}

\date{\today}

\begin{abstract}
Quanta splitting is an essential generator of Gaussian entanglement, exemplified by Einstein-Podolsky-Rosen states and apparently the most commonly occurring form of entanglement. In general, it results from the strong pumping of a nonlinear process with a highly coherent and low-noise external drive. In contrast, recent experiments involving efficient trilinear processes in trapped ions and superconducting circuits have opened the complementary possibility to test the splitting of a few thermal quanta. Stimulated by such small thermal energy, a strong degenerate trilinear coupling generates large amounts of nonclassicality, detectable by more than 3~dB of distillable quadrature squeezing. Substantial entanglement can be generated via frequent passive linear coupling to a third mode present in parallel with the trilinear coupling. This new form of entanglement, outside any Gaussian approximation, surprisingly grows with the mean number of split thermal quanta; a quality absent from Gaussian entanglement. Using distillable squeezing we shed light on this new entanglement mechanism for nonlinear bosonic systems.
\end{abstract}

\maketitle

\textit{Introduction}---
Entangled systems form the backbone of many fields of quantum physics and applications in quantum technology. Since the introduction of the fundamental Einstein-Podolsky-Rosen (EPR) states~\cite{einstein_can_1935} the most frequently occurring entanglement seems to be Gaussian entanglement. In such cases, a coherent, and low-noise external drive is used to linearise nonlinear systems and therefore the covariance matrix provides a full description~\cite{braunstein_quantum_2005}. However, it is widely anticipated that strongly nonlinear quantum interactions exhibit a much richer and exciting variety of phenomena which can potentially produce entanglement from low thermal energy, although typically at the cost of ease of detection, simple explanations, and analytical descriptions. Such novel entanglement might be much more common than anticipated and even appear more autonomously via small local thermal excitation without the need for a strong, coherent, and low-noise external drive.  

In this Letter, we analyse the deterministic and thermal creation of just such a new form of entanglement, arising from the splitting of a few thermal quanta in trilinear interactions, with the logarithmic negativity increasing with {\em increasing} thermal occupation. We use distillable squeezing instead of conventional squeezing as a new and proper diagnostic tool to analyse such hidden entanglement beyond the Gaussian approximation. We thus begin our discussion from the lowest order nonlinear interactions between linear oscillators, usually the most feasible in experiments. Nondegenerate trilinear couplings, while producing the EPR type of Gaussian entanglement under a classical pump~\cite{ou_realization_1992}, do not produce bipartite entanglement from a thermal pump among any combination of the modes~\cite{gonzalez_continuous-variable_2018,agusti_tripartite_2020} (see also Supplementary Material (SM)
). Hence, we focus on a degenerate trilinear interaction generating significant nonclassicality from splitting just a few thermal quanta. When the degenerate interaction is continuously combined with a linear coupling to an auxiliary mode the resulting enhancement of logarithmic negativity can unconditionally reach $\sim0.77$, saturating the entanglement potential~\cite{asboth_ep_prl_05}. Thermally induced entanglement (TIE) thus materialises straightforwardly via the splitting of a few thermal quanta in a regime very different to the Gaussian regime involving a strong coherent pump. To investigate the difference in these regimes we examine in detail the distribution and concentration of distillable squeezing~\cite{filip_distillation_2013} in such entangled states and compare with conventional squeezing from Gaussian entangled states.

\textit{Trilinear Interactions in Bosonic Platforms}--- The mechanical motion of trapped atoms is an excellent candidate for a proof-of-principle investigation of TIE due to several properties: (i) the strong nonlinear coupling of highly linear mechanical oscillators, (ii) low damping and substantial shielding from ambient background noises and (iii) the potential scalability of the trapped atoms~\cite{leibfried_RMP_2003,tiecke_nanophotonic_2014}. For the mechanical motion of electrically trapped ions, the intrinsic nonlinearity of the Coulomb coupling can be leveraged into several multimode versions of a trilinear Hamiltonian~\cite{nie_theory_2009} of the type historically studied to investigate generalisations of squeezing to higher order moments~\cite{braunstein_generalized_1987}. The trapped ion versions of such Hamiltonians, both partially degenerate and nondegenerate, comprising two-mode and three-mode nonlinearities respectively, have been experimentally achieved~\cite{ding_quantum_2017} and shown to be effective for tasks such as quantum simulation~\cite{ding_quantum_2018} and quantum refrigeration~\cite{maslennikov_quantum_2019}. For these reasons, we prefer to frame our discussion around systems of trapped ions. In the microwave domain, however, superconducting circuits may take advantage of the nonlinearity of Josephson junctions to generate parametric amplification~\cite{perelshtein_broadband_2021},  three-photon downconversion~\cite{chang_observation_2020} and induce trilinear Hamiltonians (SNAILs)~\cite{frattini_3-wave_2017,quijandria_universal_2020,vrajitoarea_quantum_2020} outside the rotating wave approximation. Furthermore, the TIE we describe may stimulate new nonlinear optics experiments using second- and high-harmonic generation~\cite{grosse_observation_2008,olsen_third-harmonic_2018}, three and four wave mixing~\cite{boyer_entangled_2008,coelho_three-color_2009,gonzalez_continuous-variable_2018} or multiphoton Kerr processes~\cite{silberhorn_generation_2001,gabriel_entangling_2011}. 

\textit{Thermally Induced Nonclassicality and Entanglement}--- A pair of ions with equal mass $m$ and charge $e$, contained in a harmonic trap, are coupled via the Coulomb interaction. If the radial trapping frequencies $\omega_x$ and $\omega_y$ are assumed to be much greater than the axial trapping frequency $\omega_z$, then the ions distribute themselves along the $z$-axis. The Coulomb interaction can be expanded to second order, inducing a natural transformation to the normal modes of the motion, wherein the collective vibrational modes are decoupled from each other. The first step beyond this harmonic approximation, involving third order terms in the expansion, removes the decoupling of the spatial motion leading to interactions involving both the radial and axial modes. Components of these nonlinear interactions are resonant properties of the collective motion of the ions. From here, we will assume that the most significant components of the potential are those contributing to the interaction between the $x$ and $z$ spatial components. In the rotating wave approximation, the interaction Hamiltonian takes the trilinear form
\begin{equation}
    H=\Omega_T\left(ab^\dagger{}^2+a^\dagger b^2\right),
    \label{eqn:p_ham}
\end{equation}
where $\Omega_T  =  \frac{3e^2}{4|Q_0|^4}$. For a more detailed Hamiltonian analysis see, for instance, Refs.~\cite{marquet_phononphonon_2003,lemmer_quantum_2018} or the SM. This is a partially degenerate trilinear interaction in which the creation (annihilation) of a phonon in the axial direction results in the annihilation (creation) of a pair of phonons in the radial direction.

\begin{figure}[ht!]
    \centering
    \hspace{3ex}
    \includegraphics{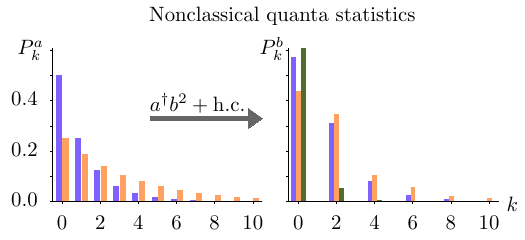}\\ \vspace{1ex}
    \includegraphics{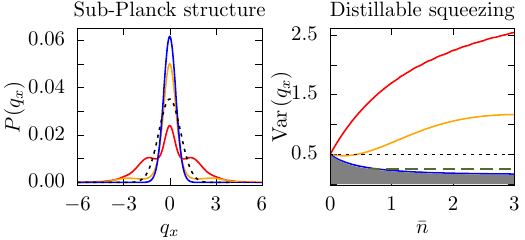}
    \vspace{-4ex}
    \caption{Nonclassical quanta statistics: The coupling of mode $a$ to mode $b$ through the trilinear coupling (Eq.~\eqref{eqn:p_ham}) dynamically splits thermal quanta in $a$ into pairs in $b$, generating phase-insensitive nonclassical phonon distributions, verified by Klyshko's criteria. Here $b$ is in the ground state (for thermal occupation of $b$, see SM). Increasing $\bar{n}$ makes the nonclassical effects more pronounced. Purple and orange correspond to $\bar{n}=1$ and 3, respectively, while green corresponds to a squeezed state with 3~dB of squeezing. Sub-Planck structure: The probability densities $P(q_x)$, of the dimensionless quadrature $\hat{q}_x=\frac{b+b^\dagger}{\sqrt{2}}$, are compared for the ground state (dashed black) and with thermally induced nonclassicality, $\bar{n}=3$, (red) demonstrating the sub-Planck structure around the origin generated by the nonlinear interaction, with the colours corresponding to the resulting distilled distributions on the right. Distillable Squeezing: The nonclassicality of mode $b$ can also be quantified in a phase sensitive manner via distillable squeezing. Universal distillation of squeezing from multiple copies of the nonclassical state in mode $b$ is shown on the lower right, with the asymptotic limit in grey~\cite{filip_distillation_2013} and the 3~dB squeezing limit in dashed green. Consuming progressively more copies produces a variance moving towards and then below the shot-noise level (dashed black). The distillation procedure saturates (blue), limited by the nonclassicality present in mode $b$ after the trilinear interaction.}
    \label{fig:sketch1}
\end{figure}

The quanta splitting and ensuing nonclassicality (necessary for entanglement) from the nonlinear dynamics may proceed from a small amount of classical thermal noise present in one of the subsystems. By initialising the axial ($a$) or radial ($b$) components of the motion in thermal states while the remainder is initialised in the ground state, the degree to which the nonlinearity converts the initial resource of thermal noise into nonclassical behaviour can be examined (Fig.~\ref{fig:sketch1}). In general, if the radial mode is used as a thermal pump of the axial mode then the nonclassicality and corresponding entanglement potential is small. In contrast, our starting point is that $a$, using the small incoherent energy of a thermal pump, can drive nonclassicality in $b$.

In the ideal case, the nonclassicality of mode $b$ can be immediately seen from the phonon distribution $P_k=\braket{k|\rho|k}$; the reduced state of mode $b$ is nonclassical due to its population of only even Fock states irrespective of the strength of the thermal pump. This is verified directly by Klyshko's criteria~\cite{klyshko_observable_1996,innocenti_nonclassicality_2022}. The nonclassicality is phase insensitive, reflecting the incoherent phase of the thermal pump (see SM for comparison with a coherent pump). Simultaneously, the position probability density, despite having a variance greater than the ground state, displays sub-Planck structure~\cite{zurek_sub-planck_2001} around the global maximum verified by the presence of universally distillable squeezing~\cite{filip_distillation_2013}, with the degree of squeezing available increasing with the thermal energy. Notably, despite the presence of such sub-Planck structures, the Wigner function is positive at the level of the nonclassical state of $b$. The impurity of such non-Gaussian states precludes detailed categorisation of the quasi-probability distribution~\cite{hudson_when_1974,soto_when_1983} and is another motivation to use distillable squeezing. The lower right panel of Fig.~\ref{fig:sketch1} shows that the degree of distillable squeezing stimulated by quanta splitting is {\it enhanced} by increases in $\bar{n}$ beyond 3~dB. The nonclassicality is remarkably robust, remaining detectable both by Klyshko's criteria and distillable squeezing despite an initial thermal occupation greater than one phonon in the quanta splitting mode (see SM).

The amount of nonclassicality capable of being converted to entanglement in a particular mode can be further characterised through the entanglement potential (EP)~\cite{asboth_ep_prl_05}, denoted  $\mathcal{E}$. The EP of mode $a$ or $b$ is evaluated through its capacity to produce logarithmic negativity (LN)~\cite{plenio_logarithmic_2005} via passive interactions with the ground state of an oscillator; fully classical states will not produce entanglement this way. The LN is calculated by determining $\mathcal{L}(\rho)=\log_2||\rho^{T_A}||$, where $T_A$ denotes the partial transpose with respect to a subsystem $A$. 
For the EP, the LN should be calculated for the following state:
\begin{equation}
 \displaystyle \rho_\mathcal{E}=U_\text{BS}^\dagger\,\left(\rho\otimes\ket{0}\bra{0}\right)U_ \text{BS}, 
 \label{eqn:bs}
\end{equation}
where $U_{\text{BS}} = e^{\frac{\pi}{4}(c^{\dagger} A - c A^{\dagger})}$, $c$ is an auxiliary mode prepared in the ground state $\ket{0}$ and $\rho$ is the reduced state corresponding to the mode $A=a,b$. EP links the nonclassicality produced by the trilinear interaction to the upper bound~\cite{vidal_computable_2002} on the entanglement generated between the nonclassical mode $b$ and an auxiliary classical mode $c$ and will allow us to directly compare the EP with the dynamically achievable TIE~\cite{zhang_non-gaussian_2021}. 

With this in mind, the nonclassicality converted from the thermal noise resource can be leveraged to create entanglement with another mode $c$ which interacts simultaneously with $b$ via the passive linear interaction
\begin{align}
    H_\text{aux} &= H + g(b\, c^{\dagger} +  b^{\dagger}c)\,.
    \label{AuxMode}
\end{align}
In our analysis, we have selected $g=\Omega_T$, which appears to give the best performance. Deviation from equality simply results in reduced LN, although the qualitative effects remain the same. The rise in LN is not a short-term effect, in the sense that it arises after an initial quiescent period in the dynamics, and optimisation of the interaction time necessarily occurs over anharmonically oscillating dynamics. To compromise we select the first peak of LN in order to balance finding significant entanglement while remaining closer to the short times available to experimental settings, where decoherence will be less relevant.

\begin{figure}[ht!]
    \centering
    \includegraphics[width=\columnwidth]{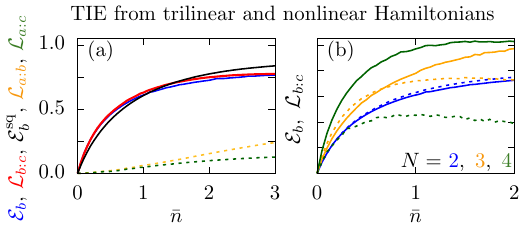}
    \vspace{-3ex}
    \caption{TIE from a compound trilinear and linear process(a) For the trilinear coupling given by Eq.~\eqref{eqn:p_ham}, the entanglement potential $\mathcal{E}_b$ (blue) of mode $b$ increases as the thermal noise is increased. After passively interacting [Eq.~(\ref{AuxMode})] with an auxiliary mode $c$, entanglement measured by logarithmic negativity $\mathcal{L}$ is shown between $b$ and $c$ (red), $a$ and $b$ (yellow), and $a$ and $c$ (green). Both $b$ and $c$ are initially in their respective ground states. The values plotted here correspond to the first peak in the temporal evolution of their respective quantities $\mathcal{E}$ and $\mathcal{L}$ which otherwise undergo rather complex evolution. $\mathcal{L}_{bc}$ closely mimics the behavior of $\mathcal{E}_b$ indicating that the entanglement potential of $b$ can be realised dynamically, despite residual entanglement among the remaining modes. $\mathcal{E}_b$ is further compared against the EP of a squeezed vacuum state (black), whose degree of squeezing is selected in accord with the asymptotic distillation of squeezing available for the corresponding trilinear state. (b) Entanglement potential $\mathcal{E}_b$ (solid curves) and $\mathcal{L}_{b:c}$ (dashed curves) generated by incoherent thermal noise in $a$ for Hamiltonians $H_\text{NL}$. The entanglement extracted by the auxiliary linear coupling ceases to closely follow the EP for $N>2$.}
    \label{fig1:ep_ln_nbar}
\end{figure}

The EP generated in $b$ by a thermal pump in $a$ is shown in Fig.~\ref{fig1:ep_ln_nbar} (left) as the blue line. If a Gaussian squeezed state is prepared with the asymptotic value of the distillable squeezing (Fig~\ref{fig:sketch1}) then this squeezed state generates similar EP (black line) to the non-Gaussian state from the trilinear interaction. The auxiliary mode $c$ simultaneously coupled to $b$, as in Eq.~(\ref{AuxMode}), results in LN, $\mathcal{L}_{b:c}$, that dynamically fulfills the EP of $b$. This can be seen as the red line in Fig.~\ref{fig1:ep_ln_nbar} (left) achieves at least the value of $\mathcal{E}_b$. This persists even though residual LN is generated with the thermal pump mode ($\mathcal{L}_{a:c}$ and $\mathcal{L}_{a:b}$) and is subsequently traced out. The Gaussian entanglement, extracted from the covariance matrix of the state, is always zero during the evolution due to the thermal pump, indicating that the thermally induced entanglement (TIE) involves correlations beyond the covariance matrix and is properly referred to as non-Gaussian entanglement.

For Gaussian entanglement either quadrature squeezing, or a generalised squeezing emerging from concentration of squeezing via linear interference, can always be used to detect the nonclassicality of the correlations. To analyse non-Gaussian entanglement beyond the covariance matrix in an operational and systematic way, backwards compatible with Gaussian entanglement and quadrature squeezing, we have relied on distillable squeezing. We have dedicated a detailed discussion in the SM to support the use of this new and extendable tool while this Letter focuses on the significant result that substantial non-Gaussian entanglement can arise from feasible and thermally driven nonlinear interactions.

To illustrate the robustness of these phenomena we examined the deviation of mode $b$ from the ideal ground state. Thermal occupation of mode $b$, while reducing the absolute value of the EP, maintains the enhancement of nonclassicality due to increasing the thermal energy in $a$. Remarkably, this occurs even when the thermal occupation of $b$ is greater than 1 and can occur when the thermal occupation of $b$ is greater than that of $a$. This allows experimental tests of these phenomena without being bound to the ground state of mode $b$. Additionally, we examined the effect on the EP and LN when the trilinear interaction occurs outside the deep-strong coupling regime, so that the free evolution of the oscillators contributes, something not present in trapped ions~\cite{marquet_phononphonon_2003}. While the EP is robust to these effects, the LN decreases, while still retaining the property that it increases with $\bar{n}$. Finally we also examined the effect of coupling the modes to a thermal environment. While both EP and LN are reduced by the interaction with the environment, significant values of both quantities remain (see SM).  

\textit{Higher Order Nonlinearities}--- It is possible that higher nonlinearities will exhibit diverse phenomena and so we examine highly nonlinear Hamiltonians of the form $H_\text{NL}\propto a(b^\dagger)^N+a^\dagger b^N$, where $N>2$. Continuing the logic, the internal thermal driving of $a$ will drive higher order quanta splitting, resulting in generalised squeezing~\cite{braunstein_generalized_1987} in $b$ and producing highly nonclassical and non-Gaussian states with large entanglement potential.

\begin{figure}[ht!]
    \centering
   \includegraphics{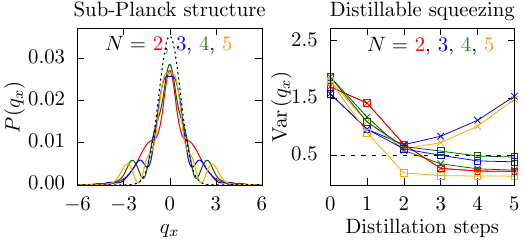}
    \vspace{-3.0ex}
    \caption{(a) The probability densities $P(q_x)$ from $H_{\text{NL}}=\Omega_T(a^\dagger b^N + ab^{\dagger\, N})$ with $N=2$ (red), 3 (blue), 4 (green) and 5 (yellow) are compared at the optimal time (see text), compared with the ground state distribution (dashed black). (b) Universal (squares) and non-universal (crosses) squeezing distillation as a function of number of copies of the state. For odd $N$ the universal method does not find distillable squeezing around the global maximum of the distribution, unlike the case of even $N$. Instead, the non-universal method searches the whole distribution for nonclassical structures and distills them to find distillable squeezing. In both panels mode $a$ is initially in a thermal state with $\bar{n}=1$ and mode $b$ is in the ground state.}
    \label{fig2:dist_sqz_all}
\end{figure}

In Fig.~\ref{fig2:dist_sqz_all} we show the probability densities in $q_x$ for the nonclassical states produced by these higher order nonlinearities. They are clearly non-Gaussian, with embedded sub-Planck structures. However, the distillable squeezing shows features distinct from the $N=2$ case considered above. The universal method of distillation~\cite{filip_distillation_2013}, which extracts nonclassical features from the curvature around the global maximum of the distribution, does not produce squeezing for odd $N$. Indeed this is reflected in the distributions which have low curvature around their maxima. Instead, the non-universal method~\cite{filip_distillation_2013,filip_squeezed-state_2014}, which optimises the distillation around arbitrary features of the distribution, can extract squeezing even from the odd parity Hamiltonians.
However, the distillable squeezing found through the non-universal method does not increase with $\bar{n}$, in contrast with the universal method analysed in detail in Fig~\ref{fig:sketch1}.

This nonclassicality can again be used to generate entanglement, again using the simultaneous linear coupling as used in Eq.~(\ref{AuxMode}). Fig.~\ref{fig1:ep_ln_nbar} (right) shows the entanglement potential and logarithmic negativity via coupling to an auxiliary state for various $N$. Increasing $N$ tends to increase the nonclassicality generated by the same thermal energy, however, the anticipated generation of entanglement fails to reach the EP for larger values of $\bar{n}$. This suggests that among these the trilinear Hamiltonian of Eq.~(\ref{eqn:p_ham}), combined with a linear coupling to another mode, is the most effective in terms of generating non-Gaussian entanglement from small thermal energy.

\textit{Discussion}--- We have singled out the simplest nonlinearity coupling linear oscillators, the degenerate trilinear interaction, combined with a linear coupling in order to transmute the thermal energy of a few quanta into nonclassicality and non-Gaussian TIE, taking us simultaneously beyond the more common Gaussian EPR-type states and non-Gaussian pure states understood by Hudson's theorem~\cite{hudson_when_1974,soto_when_1983}. Remarkably the quantum non-Gaussianity manifests differently and positively as entanglement which grows as the thermal energy is increased. Other possibilities among three mode interactions are surprisingly limited in their capacity to generate bipartite entanglement~\cite{gonzalez_continuous-variable_2018,agusti_tripartite_2020} (see SM). That larger thermal fluctuations are converted into larger coherent quantum phenomena, such as entanglement~\cite{bose_subsystem_2001,kim_entanglement_2002,marek_deterministic_2016,laha_thermally_2021}, functions as a direct proof of genuine nonlinear quantum effects, as it cannot happen for linearised dynamics within the Gaussian approximation provided by the covariance matrix. While such dynamics intrinsically involves higher than quadratic interaction terms, and shows quantum non-Gaussian effects~\cite{ding_quantum_2017} such a direct experimental proof is yet to be demonstrated. In the past, such features were predicted in systems taking advantage of discrete nonlinearities in the level structure of atoms or superconducting circuits with very strong saturation~\cite{marek_deterministic_2016,slodicka_deterministic_2016,laha_thermally_2021} while here entanglement materialises without the direct use of saturation in such discrete nonlinearities. It is also conceptually different to TIE using controlled-SWAP gates~\cite{filip_bell-inequality_2002,gao_entanglement_2019}. The thermally induced nonclassicality and entanglement studied here arises purely from the natural Coulomb force between trapped ions. Analogous effects from the direct coupling, without any assisting drive, for fully discrete variable systems remains to be investigated.  

This spearhead analysis, inspired by recent experimental achievements regarding trilinear interactions in trapped ion~\cite{ding_quantum_2017,ding_quantum_2018,maslennikov_quantum_2019} and superconducting circuit platforms~\cite{chang_observation_2020,vrajitoarea_quantum_2020}, provides a clear example of the hidden power of multimode nonlinear systems to achieve, autonomously and unconditionally, increasing nonclassical quantum behaviour leveraging only small thermal energy. Furthermore, the linear coupling required to directly test the parallel generation of entanglement has already been proposed for optical systems~\cite{krastanov_room-temperature_2021} and is likely to be easily adapted to phononic systems and superconducting circuits.   

\begin{acknowledgements}
  The authors acknowledge funding from Project 21-13265X of the Czech Science Foundation. This project has received funding from the European Union's 2020 research and innovation programme (CSA Coordination and support action, H2020-WIDESPREAD-2020-5) under grant agreement No. 951737 (NONGAUSS).  
\end{acknowledgements}

\bibliographystyle{apsrev4-2}
\bibliography{references} 

\clearpage
\begin{widetext}
\begin{center}
    \textbf{\large Supplemental Material for ``Entanglement growth via splitting of a few thermal quanta''}
\end{center}
\end{widetext}

\setcounter{section}{0}
\setcounter{figure}{0}
\setcounter{table}{0}
\makeatletter
\renewcommand{\thesection}{S\arabic{section}}
\renewcommand{\thefigure}{S\arabic{figure}}
\renewcommand{\thetable}{S\arabic{table}}
\renewcommand{\theequation}{S\arabic{equation}}
\makeatother

\subsection{Evaluating non-Gaussian Entanglement by Distillable Squeezing}

\subsubsection{Distillable Squeezing as an Efficient Quantifier for non-Gaussian States}

Understanding thermally-powered non-Gaussian entanglement arising from a nonlinear interaction involving a few quanta requires a new theoretical development going beyond the covariance matrix formalism. In order to be able to compare and interpret the diversity of such non-Gaussian states, we must identify and analyse nonclassical features that are present across a wide range of non-Gaussian states, and that fulfil the basic requirements of accessibility to direct measurement and straightforward operational interpretation. We demonstrate here that the non-Gaussian nonclassical properties produced by the nonlinear interaction in the main text can be efficiently evaluated using the new concept of distillable squeezing~\cite{filip_distillation_2013}. It is an operational quantification of the sub-Planck structures~\cite{zurek_sub-planck_2001} present in non-Gaussian states which have already proved to be essential ingredients for advanced quantum sensing~\cite{wang_heisenberg-limited_2019,mccormick_quantum-enhanced_2019} and error correction~\cite{fluhmann_encoding_2019,campagne-ibarcq_quantum_2020}. It is an operationally generalised version of conventional squeezing obtained from the quadrature variance in the Gaussian approximation to quantum non-Gaussian states, including phase insensitive ones. In the Gaussian approximation, both conventional squeezing and entanglement~\cite{laurat_entanglement_2005} are captured fully through the covariance matrix. However, squeezing and entanglement of non-Gaussian states are typically not visible from the covariance matrix, and our approach is a step towards a future systematic characterization that has been until now untenable. 

Following the historical and fundamental development of conventional squeezing and entanglement analysis in the Gaussian approximation~\cite{einstein_can_1935,reid_demonstration_1989,simon_quantum-noise_1994,giedke_characterization_2002,reid_colloquium_2009}, we focus our analysis on key comparisons with paradigmatic examples of Gaussian and basic non-Gaussian entangled states and their properties in terms of conventional and distillable squeezing. Using this approach we show how, in the absence of conventional squeezing, distillable squeezing is a suitable substitute for non-Gaussian states. Distillable squeezing thus makes apparent features that are otherwise undetectable via conventional squeezing and importantly is a quantity that is both operationally defined in a universal way for an arbitrary state and commensurable with conventional squeezing. Here, naturally, we can only provide the necessary supplementary information supporting the main text, and a comprehensive methodology will be provided in the future.

\subsubsection{Evaluating non-Gaussian Entanglement via Generalised Distillable Squeezing}

\begin{figure}[h]
    \centering
    \includegraphics[width=\columnwidth]{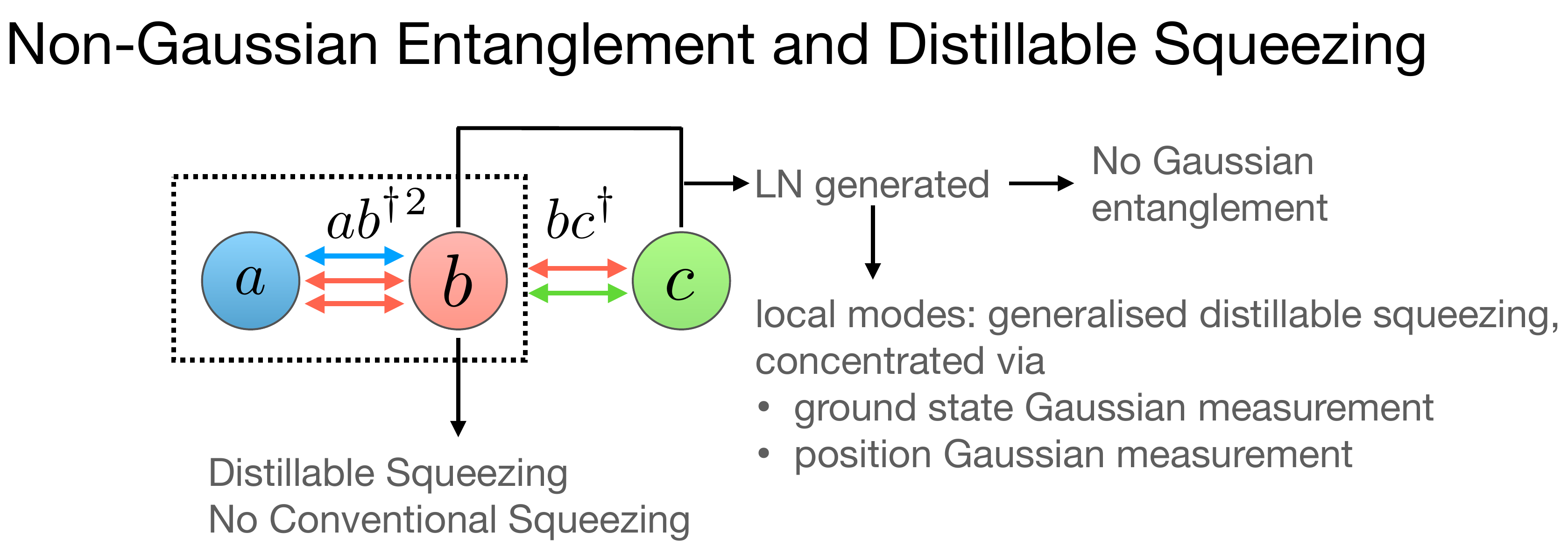}
    \caption{Analysis of non-Gaussian entanglement from the trilinear process using distillable squeezing instead of conventional squeezing.}
    \label{MainResult}
\end{figure}

First we recapitulate the main result with respect to conventional and distillable squeezing (see Fig.~\ref{MainResult}). The trilinear interaction, driven by a fully quantum thermal pump, produces in the mode $b$ (without the coupling to $c$) phase insensitive states diagonal in the Fock basis whose nonclassicality can be detected with distillable squeezing, but which show no conventional squeezing. Still the conventional squeezing might be hidden in the Gaussian approximation. In some cases, sub-Planck structures~\cite{zurek_sub-planck_2001} could be a good witness of nontrivial non-Gaussian states. Importantly, conventional squeezing can be distilled using only beamsplitter-like couplings and linear measurements, which themselves do not produce conventional squeezing. 

In the main text the entangled state produced from a further (and simultaneous) linear coupling to a third mode $c$ shows no entanglement in the Gaussian approximation and each individual mode shows neither conventional nor distillable squeezing (see column 1 of Table~\ref{tab:table2}). Distillable squeezing from mode $b$ is therefore transmuted into non-Gaussian entanglement, rather than remaining in the local states of both $b$ and $c$.  

The schematic process depicted in Fig.~\ref{MainResult} suggests an intriguing comparison of the state in the main text to two very different elementary categories of entangled states: i) Gaussian entangled states and ii) non-Gaussian entangled states obtained by splitting or interfering states squeezed in the number of quanta (Fock states). We unpack each of these broad categories in turn, using paradigmatic examples to demonstrate the diagnostic advantages of distillable squeezing.

\begin{figure}[ht]
    \centering
    \includegraphics[width=\columnwidth]{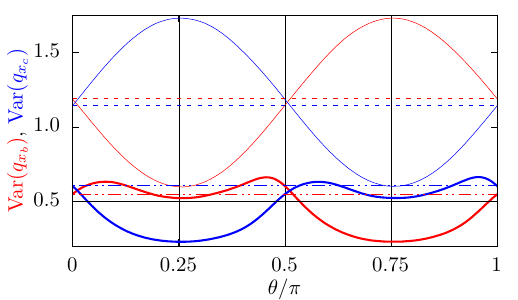}\\
    \includegraphics[width=\columnwidth]{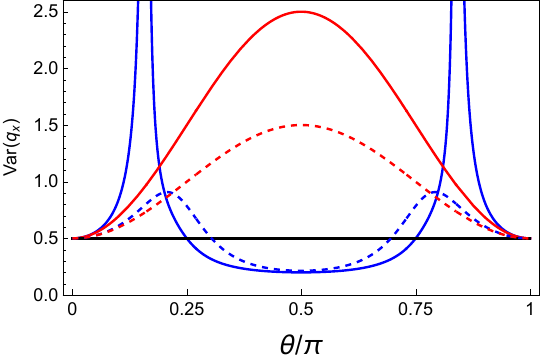}
    \caption{(Top) Generalized conventional and distillable squeezing for the trilinear interaction. The variance (thin lines) of the dimensionless quadrature $q_x$ for modes $b$ or $c$ of the non-Gaussian entangled state produced with $\bar{n}=1$, as well as the distillable variance (thick lines), as a function of $\theta$ (Eq~\ref{BS}).
    (Bottom) The variance (red) and distillable variance (blue) of the dimensionless quadrature $q_x$ for the one mode of $\rho_\text{QNG}$ (solid) and $\rho_\text{MIX}$ (dashed) as a function of as a function of $\theta$ (Eq~\ref{BS}). No generalised conventional squeezing is possible. Instead, it requires distillable squeezing to see the nonclassicality of the quantum non-Gaussian state. The black line denotes the ground state threshold.}
    \label{fig:NonGaussEntDSTrilinear}
\end{figure}

\begin{table}[ht]
\caption{\label{tab:table2}%
Local vs generalized conventional and distillable squeezing for trilinear interaction. Variances in $q_x$ for local modes $b$ and $c$ of Eq.~(6) first prepared at the optimal time with mode $a$ initially in a thermal state with $\bar{n}=1,\, 2$ (first two rows and the bottom two rows, respectively), and modes $b$ and $c$ initially in ground states. Rows below contain the variances in $q_x$ for the reduced state of either the Fock state $\ket{2}$ split with the ground state on a balanced beamsplitter ($\theta=\frac{\pi}{4}$), see Eq.~(\ref{Fock2split}), $\rho_\text{QNG}$, or the corresponding mixed state $\rho_\text{MIX}$. For the former, the generalised distillable squeezing achieved via beamsplitter operations simply corresponds to the distillable squeezing of the pure state $\ket{2}$. The right column displays the minimum variance achievable via mixing $b$ and $c$ via interaction (\ref{BS}). Bold numbers correspond to the distillable squeezing.}
\begin{ruledtabular}
\begin{tabular}{ccccc}
Mode &  Local Squeezing & Generalised Squeezing \\
\colrule
\multirow{2}{2em}{$b$} 
& 1.187 &  0.598 \\ 
& {\bf 0.548} & {\bf 0.229} \\  
\colrule
\multirow{2}{2em}{$c$} 
& 1.141 & 1.729  \\ 
& {\bf 0.603} & {\bf 0.229} \\ 
\colrule
\multirow{2}{2em}{$b$} 
& 1.537 &  0.788 \\ 
& {\bf 0.567} & {\bf 0.187} \\ 
\colrule
\multirow{2}{2em}{$c$} 
& 1.611 & 2.375  \\ 
& {\bf 0.521} & {\bf 0.187} \\ 
\hline \hline
\colrule
 \multirow{2}{2em}{$\rho_\text{QNG}(2)$} & 
1.5 &  0.5\\ 
 & 
{\bf 0.5} & {\bf 0.2}\\ 
 \colrule
 \multirow{2}{2em}{$\rho_\text{MIX}(2)$}
 & 1 &  0.5 \\ 
 & {\bf 0.786} & {\bf 0.214} \\ 
 \colrule
 \multirow{2}{2em}{$\rho_\text{QNG}(4)$} & 
2.5 &  0.5\\ 
 & 
{\bf 1.057} & {\bf 0.158}\\  
 \colrule
 \multirow{2}{2em}{$\rho_\text{MIX}(4)$}
 & 1.5 &  0.5 \\ 
 & {\bf 0.663} & {\bf 0.157} \\ 
\end{tabular}
\end{ruledtabular}
\end{table}

For two-mode Gaussian states, the Bloch-Messiah decomposition indicates that they may always be decomposed into linearly interfering single-mode squeezers. This means that the linear interference after the squeezers can always be undone so that the local squeezing operations are visible in conventional squeezing. Maximal squeezing in one of these modes is a generalized squeezing~\cite{simon_quantum-noise_1994} that can be detected using a general beamsplitter interaction on the output modes. If insufficient, a measurement can be used to demask squeezing from the classical correlations~\cite{filip_squeezing_2010}. This framework of coherently concentrating multimode features into a single mode is an inspiring and feasible path to also evaluate non-Gaussian multimode states.  

In Fig.~\ref{fig:NonGaussEntDSTrilinear}, we show the effect of attempting a similar concentration of the conventional squeezing on the modes of the non-Gaussian entangled state (see also the rightmost column of Table~\ref{tab:table2}) using the beamsplitter-like unitary operator
\begin{equation}\label{BS}
    U_\text{BS}=e^{i\theta(b^\dagger c+bc^\dagger)}\,.
\end{equation}
Note that both modes (thin red/blue lines) show zero conventional squeezing initially, and never pass below the threshold under any beamsplitter transformation. Therefore, there is no generalized conventional squeezing in the Gaussian approximation for our non-Gaussian state. Moreover, there is no distillable squeezing in the individual modes for the maximum logarithmic negativity resulting from the combined trilinear and linear interactions. The generalised distillable squeezing is however detectable by an optimal coupling (Eq.~\ref{BS}) in either mode (that is, with $\theta=\frac{\pi}{4}$ or $\theta=\frac{3\pi}{4}$). Consequently, we can observe generalised distillable squeezing for our non-Gaussian state, in analogy to generalised conventional squeezing for Gaussian states. The asymmetrical presence of the generalized squeezing in Fig.~\ref{fig:NonGaussEntDSTrilinear} shows that our state is a non-Gaussian analogy of the asymmetrical Gaussian state~\cite{takahashi_entanglement_2010,kiesel_entangled_2011,handchen_observation_2012,eberle_gaussian_2013}, where the single-mode squeezed states are split by a beamsplitter-like interaction. Unlike pure asymmetric Gaussian states that always show conventional squeezing in the individual modes~\cite{perina_quantum_1995}, the mixed state from the joint trilinear and linear coupling does not produce distillable squeezing individually. Note that while the distillable squeezing is periodic under variation of the beamsplitter gain, it loses the harmonicity of the conventional squeezing.

\subsubsection{Comparison with a split two quanta state}

The simplest paradigmatic examples of non-Gaussian entangled states to compare with the previous section are the Fock states mixed on a beamsplitter with the ground state. The simplest and closest example to our non-Gaussian entangled state, formed from an even parity state mixed with the ground state, uses the pure Fock state $\ket{2}$. This corresponds to the trilinear interaction pumped by the noiseless Fock state $|1\rangle$ for an ideal conversion time and subsequently split to two modes, which gives a squeezing upper bound for the thermal pump with $\bar{n}=1$ used in Fig.~\ref{fig:NonGaussEntDSTrilinear} and Table~\ref{tab:table2}. In general splitting the Fock state $\ket{n}$ on a beamsplitter results in the density operator
\begin{widetext}
\begin{equation}\label{Fock2split}
    \rho_\text{QNG}(n)=\frac{1}{n!}\sum_{kj}\binom{n}{k}\binom{n}{j}\sqrt{k!j!(n-k)!(n-j)!}(\cos\theta)^{k+j}(\sin\theta)^{2n-k-j}\ket{k,n-k}\bra{n-j,j}\,,
\end{equation}
\end{widetext}
where we write the diagonal elements in the notation $\ket{nm}=\ket{n}\otimes\ket{m}$.
Fixing $n=2$, this state is an upper limit to that produced by the trilinear interaction pumped by a single quantum, as the result of that process is to produce two quanta which are subsequently split, creating entanglement not described by the covariance matrix. Importantly, the reduced state is symmetric with respect to which mode is traced out, apart from a shift in $\theta$.

For a symmetrical splitting with $\theta=\frac{\pi}{4}$, this state shows no conventional squeezing and, like the state from the trilinear interaction, squeezing also cannot be universally distilled from the individual modes (see Table~\ref{tab:table2}). This happens despite the overall state being pure; therefore, it makes a substantial difference to pure squeezed states split into two. Moreover, also like the trilinear state, it is still possible to retrieve generalised distillable squeezing via general beamsplitter operations. In fact this is easy to see, as the preparation of $\varrho_\text{QNG}$ is simply undone resulting in the pure Fock state $\ket{2}$, which shows universally distillable squeezing. In contrast, the optimal conventional squeezing corresponds to the case of a total swap to the ground state, so that in this case the demarcation between extremes of Gaussian and non-Gaussian are very clear. Similar to our non-Gaussian entangled state, a general beamsplitter does not reveal conventional squeezing, but does reveal distillable squeezing. Comparing generalized distillable squeezing for the upper bound by subsequently split Fock state and thermally pumped trilinear process with simultaneous splitting, we successfully reach more than 87\% of that upper bound. To analogically compare with the $\bar{n}=2$ used in Fig.S2 and Table 1, we consider trilinear interaction pumped by the noiseless Fock state $|2\rangle$ producing optimally the Fock state $|4\rangle$ and subsequent splitting to two modes. It gives the upper bound for the distillable squeezing with the thermal pump having $\bar{n}=2$ that can be reached with a very satisfactory 84\%. Furthermore, the distillable squeezing is again anharmonic under variation of the beamsplitter gain shown in Fig.~\ref{fig:NonGaussEntDSTrilinear}. The nonclassical aspect of the non-Gaussian entanglement generation is related to the distillable squeezing in a similar way to how the nonclassical aspect of entanglement generation in the Gaussian case is related to conventional quadrature squeezing. 

Let’s now illustrate how significant the impact of fluctuations is in the simplest case of an incoherent mixture of two quanta with ground state, that is the state $\frac12\left(\ket{0}\bra{0}+\ket{2}\bra{2}\right)$. In a balanced mixture case, the Fano factor $F=\langle n^2\rangle/\langle n\rangle - \langle n\rangle = 1$ and the state has already lost sub-Poissonian statistics. The state is super-Poissonian for a more significant admixture of the ground state. This example shows that quantification through sub-Poissonian statistics is not helpful for thermally induced nonlinear phenomena. The non-Gaussian entangled state resulting from splitting on a balanced beamsplitter is modified to
\begin{equation}
    \rho_\text{MIX}=\frac12\left(\ket{0}\bra{0}^{\otimes2}+\rho_\text{QNG}\right)\,.
\end{equation}
Again with $\theta=\frac{\pi}{4}$, this has no conventional squeezing locally, local sub-Poissonian statistics do not appear and also cannot be concentrated as they are not present in input state of the beamsplitter. Finally, local distillable squeezing is also not present (see Table~\ref{tab:table2} and Fig.~\ref{fig:NonGaussEntDSTrilinear}). The mixture increases the locally distillable variance compared to $\rho_{QNG}$ generated from a split pure Fock state $\ket{2}$, even above the variance of the corresponding trilinear case. However, generalised distillable squeezing is only slightly reduced by this mixture with the ground state, which typically destroys conventional squeezing very quickly. This is an advantageous property for the detection of thermally induced quantum non-Gaussian effects. Moreover, this example shows how a mixture of Fock states quickly reduces the discussed upper bound that dropped by 50\% towards the thermally pumped trilinear process.

\begin{figure}
    \centering
    \includegraphics[width=\columnwidth]{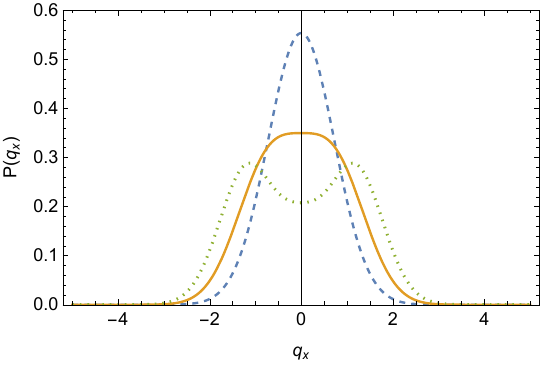}
    \caption{The probability distribution of $q_x$ for the state $\rho_\text{QNG}$ for $\theta=0.5$ (solid), $\theta=0.1$ (dashed) and $\theta=0.8$ (dotted). The incoherent mixture between $\ket{0}$ and $\ket{2}$ results in a probability distribution whose maximum vacillates between a single global maximum at $q_x=0$ associated with the ground state and a pair of displaced maxima associated with $\ket{2}$. At the transition point, the curvature is zero.}
    \label{rhoQNGtheta}
\end{figure}

It is worth briefly examining in further detail some of the limits on the bottom panel of Fig~\ref{fig:NonGaussEntDSTrilinear} as they pertain to using distillable squeezing as a tool for understanding non-Gaussian states, although once more we stress that we cannot give a full investigation here. Notably at $\theta\approx0.15$ distillable squeezing of $\rho_\text{QNG}$ diverges. In fact, this is due to the fact that the universal distillable squeezing (an asymptotic quantity) depends on the curvature of the probability distribution around the global maximum~\cite{filip_distillation_2013}. We show a few of these distributions for various $\theta$ in Fig~\ref{rhoQNGtheta}. As mentioned, the generalized distillable squeezing is very close for both the pure and mixed versions of the non-Gaussian entangled state, which can also be seen in Table~\ref{tab:table2}, despite the very strong mixing with the ground state. This occurs at $\theta=\frac{\pi}{2}$, so that $\rho_\text{QNG}=\ket{2}$ and $\rho_\text{MIX}=\frac12\left(\ket{0}\bra{0}+\ket{2}\bra{2}\right)$. 

\subsubsection{Single Quantum Driven Trilinear Interactions}

The non-Gaussian entangled state from the main text is produced from a trilinear interaction fueled by thermal noise and is analysed in terms of distillable squeezing, which we justify above. Here we analyse the trilinear interaction in the case $\bar{n}=1$ but without phonon noise, which corresponds to a trilinear process driven by the single phonon state $\ket{1}$. Using the Hamiltonian in Eq.~(4) and the initial state $\ket{10}$, the Schr\"{o}dinger equation for the trilinear process can be solved analytically by the ansatz $\ket{\psi(t)}=\alpha(t)\ket{10}+\beta(t)\ket{02}$, where 
\begin{align}
    \alpha(t)&=\cos\left(\sqrt{2}gt\right)\\
    \beta(t)&=i\sin\left(\sqrt{2}gt\right)\,.
\end{align}
Setting $g=1$, the conventional and distillable squeezing of mode $b$ can be evaluated as a function of time. In fact, the reduced state corresponding to mode $b$ oscillates incoherently between $\ket{0}$ and $\ket{2}$. 
\begin{figure}
    \centering
    \includegraphics[width=\columnwidth]{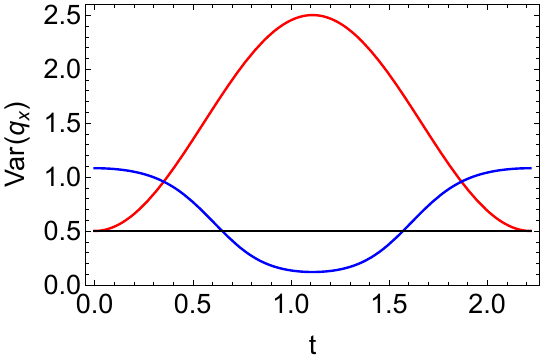}
    \caption{The variance of $q_x$ as a function of time for a trilinear interaction (Eq.~4) driven by a single phonon. The conventional squeezing (red) never goes below the value of the ground state (black), while the distillable squeezing (blue) detects sub-Planck structure even a single distillation step and filtration. The Gaussian filter is not optimised at each point.}
    \label{QuantaTri}
\end{figure}
As the state oscillates incoherently between zero and two phonons, the probability distribution oscillates between one with two global maxima symmetric about the origin (due to $\ket{2}$) and a single maximum at $x=0$ (due to $\ket{0}$). This introduces a difficulty for the asymptotic universal distillation procedure relied on so far. However the method of distillable squeezing can be adapted; if the input state is split on a beamsplitter of transmission strength $T$ and homodyne detection is performed on one arm squeezing can be distilled via an induced filtering with a Gaussian function~\cite{filip_distillation_2013,filip_squeezed-state_2014}. More generally, a finite rather than asmyptotic distillation process can be followed up with this filter in order to maximise the possible distillation squeezing. After consuming two copies (one distillation step), squeezing is already visible using this process. As shown in Fig.~\ref{QuantaTri} the conventional squeezing never goes below the value of the ground state, while for certain times distillable squeezing can be achieved, as could be seen with the case of the Eq.~4 driven by thermal noise. Note that the Gaussian filter is still not optimised each time $t$.

\begin{figure}
    \centering
    \includegraphics[width=\columnwidth]{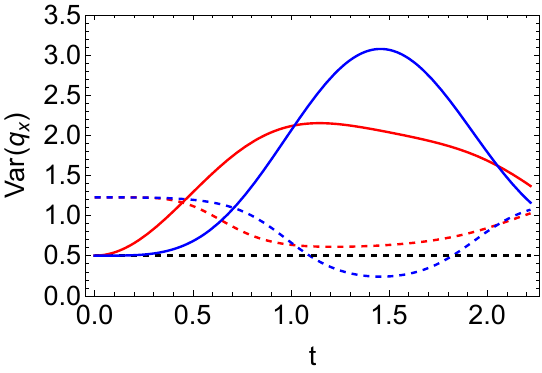}
    \caption{The variance of $q_x$ for modes $b$ (red) and $c$ (blue) as a function of time for a trilinear interaction (Eq.~6) driven by a single phonon. The conventional squeezing (solid) never goes below the value of the ground state (black), while the distillable squeezing (dashed) detects nonclassicality. }
    \label{QuantaTriLin}
\end{figure}

When a linear coupling to a third mode $c$ is introduced (Eq.~6) the Schr\"{o}dinger equation can be solved by the ansatz $\ket{\psi(t)}=\alpha(t)\ket{100}+\beta(t)\ket{020}+\gamma(t)\ket{011}+\delta(t)\ket{002}$. This is solved numerically and the squeezing properties are displayed in Fig.~\ref{QuantaTriLin}. As with an initial thermal state, the reduced states of modes $b$ and $c$ do not show conventional squeezing. In addition, as with the idealised examples of entanglement above, distillable squeezing is difficult to find locally. After the consumption of three copies (two distillation steps) and filtration only one mode begins to show distillable squeezing. Instead, a method implementing generalised distillable squeezing is required. 

\subsubsection{Evaluating non-Gaussian Entanglement via Measurement-Induced Distillable Squeezing}

\begin{table}[ht!]
\caption{\label{tab:table3}%
Variances in $q_x$ for modes $b$ and $c$ of Eq.~(6) first prepared at the optimal time with mode $a$ initially in a thermal state with $\bar{n}=1,\, 2$ (first two rows and the bottom two rows, respectively), and modes $b$ and $c$ initially in ground states, then measured, either via the homodyne detection of $q_x$ or projection onto the ground state $\ket{0}\bra{0}$, which has variance 0.5. Bold numbers correspond to the distillable squeezing. Variances in $q_x$ for the reduced state of the Fock state $\ket{2}$ split with the ground state on a balanced beamsplitter, see Eq.~(\ref{Fock2split}) and the corresponding mixed state $\frac12\left(\ket{0}\bra{0}+\ket{2}\bra{2}\right)$ after projection onto the Gaussian states $\ket{0}\bra{0}$ and $\ket{x=0}\bra{x=0}$.}
\begin{ruledtabular}
\begin{tabular}{ccccc}
Mode &  Local Squeezing & $\ket{x=0}\bra{x=0}$ & $\ket{0}\bra{0}$ \\
\colrule
\multirow{2}{2em}{$b$} 
& 1.187 & 1.208 & 0.648 \\ 
& {\bf 0.548} & {\bf 0.429} & {\bf 0.340} \\  
\colrule
\multirow{2}{2em}{$c$} 
& 1.141 & 1.121 & 0.590 \\ 
& {\bf 0.603} & {\bf 0.455} & {\bf 0.356} \\ 
\colrule
\multirow{2}{2em}{$b$} 
& 1.537 & 1.600 & 0.665 \\  
& {\bf 0.567} & {\bf 0.450} & {\bf 0.287} \\ 
\colrule
\multirow{2}{2em}{$c$} 
& 1.611 & 1.65 & 0.796 \\ 
& {\bf 0.521} & {\bf 0.3} & {\bf 0.259} \\ 
\colrule
\multirow{2}{2em}{$\rho_\text{QNG}$} & 
1.5 & 1.167 &  2.5\\ 
 & 
{\bf 0.5} & {\bf 0.167} & {\bf 0.2}\\   
 \colrule
 \multirow{2}{2em}{$\rho_\text{MIX}$}
& 1 & 0.971 &  0.9 \\ 
& {\bf 0.786} & {\bf 0.198} & {\bf 0.346} \\ 
\end{tabular}
\end{ruledtabular}
\end{table}

Concentration of two-mode generalized distillable squeezing can also be done via Gaussian measurements, classical communication, and local classical operations, without using the beamsplitter interaction as in the previous section to obtain generalized squeezing. That is, sometimes squeezing can be concentrated in one mode of an unknown two-mode asymmetrical Gaussian entangled state via projection onto either the Gaussian ground state, shunting all energy into the remaining mode, or more efficiently by optimised homodyne detection. Here we take this to be a measurement of the position quadrature $X$ conditioned on $x=0$, as the distillable squeezing is also reached conditionally. Table~\ref{tab:table3} shows the results of this for our non-Gaussian entangled state. No conventional squeezing results from either measurement, however, they both can be used to steer the detectable distillable squeezing. Naturally, the measurement-induced distillable squeezing is smaller than the generalized distillable squeezing, in analogy with Gaussian measurement-induced and generalized squeezing~\cite{filip_squeezing_2010}. Phase insensitive projection on the ground state results in greater distillable squeezing than the phase sensitive homodyne detection, which is never the case for Gaussian states and conventional squeezing. 

The paradigmatic example of pure non-Gaussian entanglement, the result of splitting the Fock state $\ket{2}$ ($\rho_\text{QNG}$ with $\theta=\frac{\pi}{2}$), also shows its own squeezing properties under measurement. A projection onto the ground state leaves the remaining mode in the Fock state $\ket{2}$, which has no conventional squeezing, yet has strong distillable squeezing, as noted above (Table~\ref{tab:table3}). For such states, projection on the ground state is in fact equivalent to concentration by linear interference. For the mixed version, $\rho_\text{MIX}$, projection on the ground state does not retrieve the original mixed state, but rather a modified state
\begin{equation}
    \rho_{\text{MIX},\ket{0}\bra{0}}=\frac45\left(\ket{0}\bra{0}+\frac14\ket{2}\bra{2}\right)\,.
\end{equation}
This state has a greater conventional squeezing, but the variance does not reduce below that of the ground state. The distillable squeezing does reveal the nonclassicality but is reduced in comparison with $\rho_\text{QNG}$.

Projecting $\rho_\text{QNG}$ on to the position eigenstate with $x=0$, results in 
\begin{equation}
\rho_{QNG,X}=\frac23\left(\frac12\ket{0}\bra{0}-\frac{1}{\sqrt{2}}(\ket{0}\bra{2}+\ket{2}\bra{0})+\ket{2}\bra{2}\right)\,.
\end{equation}
This conditional state also has no conventional squeezing, but shows a high degree of distillable squeezing (Table~\ref{tab:table3}) without concentration via linear interference as in the previous section. However, projecting on the ground state is less effective than homodyne detection, unlike our non-Gaussian entangled state and similar to Gaussian states. If the diagonal elements are removed, the distillable squeezing still takes the value $0.167$. For the mixture $\rho_\text{MIX}$, the same homodyne projection results in
\begin{equation}
\rho_{MIX,X}=\frac{8}{17}\left(\frac98\ket{0}\bra{0}-\frac{1}{\sqrt{2}}(\ket{0}\bra{2}+\ket{2}\bra{0})+\ket{2}\bra{2}\right)\,.
\end{equation}
This state has lower conventional squeezing compared with $\rho_{QNG,X}$, but again does not go below the ground state value. However at the level of distillable squeezing the reduction in variance is comparable to, although less than, that of $\rho_{QNG,X}$. Again homodyne detection concentrates the distillable squeezing more effectively than projection on the ground state. This is more reflective of Gaussian behaviour than the non-Gaussian behaviour seen in our non-Gaussian entangled state. When the diagonal elements are removed the distillable squeezing is decreased to $0.224$.

\subsection{Nondegenerate and Multiple Trilinear Hamiltonians}
\begin{figure}[ht!]
    \centering
    \includegraphics{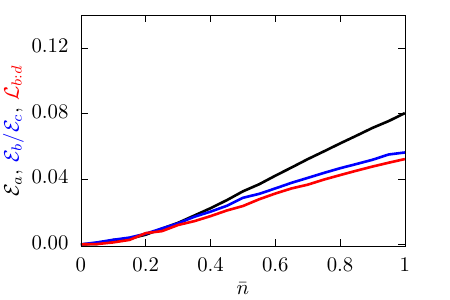}
    \vspace{-1ex}
    \caption{EP of mode $a$ (black), and identical modes $b$ and $c$ (blue) are plotted as a function of $\bar{n}$ for the nondegenerate trilinear system given in Eq.~\eqref{eqn:nond_ham}. $a$ is prepared in a thermal state characterised by $\bar{n}$ while $b$ and $c$ are in the ground state. Adding a linear coupling between mode $b$ and an auxiliary mode $d$, as in $H^{\mathrm{ND}}_{\textrm{aux}}$, the LN $\mathcal{L}_{b:d}$ increases with $\bar{n}$ but is negligible compared to the degenerate case.}
    \label{fig:ep_a_b_nbar_nond}
\end{figure}

In the main analysis, we focused on degenerate trilinear coupling as the best candidate for the generic entanglement generation from a few thermal quanta. Alternative configurations of trilinear Hamiltonians are briefly analysed here and demonstrated to be inefficient in producing entanglement. In the nondegenerate case, the Hamiltonian is given by
\begin{equation}
    H^{\mathrm{ND}} = \Omega (a^{\dagger} \,b\,c + a\, b^{\dagger}\, c^{\dagger}),
    \label{eqn:nond_ham}
\end{equation}
where $\Omega$ is the coupling strength. This interaction Hamiltonian has already been realised using three ${}^{171}\mathrm{Yb}^+$ ions in a Paul trap,  where $a$ corresponds to the axial mode, and $b$ and $c$ to the two radial modes~\cite{ding_quantum_2018}. Thermally pumping any one of the modes $a,\,b,\,c$, does not produce entanglement among the remaining pair. However, we can check the EP of these modes individually and extract the entanglement via continuous linear coupling to an auxiliary mode $d$. Setting $\tau =  \Omega t$ as before, and studying the induced nonclassicality in mode $b$ (equivalent to $c$ by symmetry) by thermally driving mode $a$, while both modes $b$ and $c$ are initially in their respective ground states we find (Fig.~\ref{fig:ep_a_b_nbar_nond}) that EP can be generated in mode $b$ ($c$), in negligible quantities compared to the degenerate case analysed in the main text. As before, we also see that the nonclassicality converted from the thermal noise resource can be leveraged via coupling $b$ to an auxiliary mode $d$ (i.e. $ H^{\mathrm{ND}}_{\textrm{aux}} = H^{\mathrm{ND}} + g(bd^\dagger+b^\dagger d)$, or similarly with $c$) to create entanglement which follows the behavior of $\mathcal{E}_{b}$ ($\mathcal{E}_{c}$). Again we set $g=\Omega$. We note that $\mathcal{E}_a$ is greater than $\mathcal{E}_b$ (or equivalently, $\mathcal{E}_c$) which is not the case for the degenerate case investigated in the main text. 

\begin{figure}[ht!]
     \centering
     \includegraphics{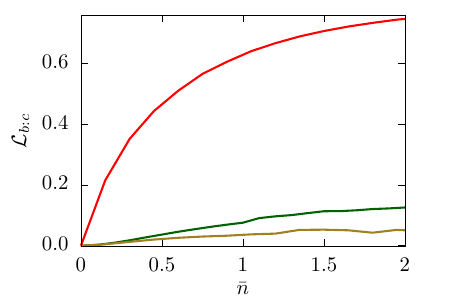}
     \vspace{-1.0ex}
     \caption{$\mathcal{L}_{b:c}$ for $H_{1}$ (olive),  $H_{2}$ (green) is compared with $\mathcal{L}_{b:c}$ for $H$  (red) in Fig.~2
     . The LN for multiple trilinear interactions is much smaller than for the single degenerate version preferred in the main text.}
     \label{fig:trip_2sq_ln_comp}
 \end{figure}

Alternatively, multiple trilinear interactions with one common mode are described by the following Hamiltonians, 
\begin{align}
    H_1 &= \Omega_1(ab^\dagger{}^2+a^\dagger b^2)+\Omega_2(cb^\dagger{}^2+c^\dagger b^2)\\
    H_2 &= \Omega_1(ab^\dagger{}^2+a^\dagger b^2) + \Omega_2 (ac^\dagger{}^2+a^\dagger c^2).
\end{align}
In Fig.~\ref{fig:trip_2sq_ln_comp} we show the behavior of $\mathcal{L}_{b:c}$ for $H_1$ (olive) and $H_2$ (green). In both the cases the induced entanglement is much smaller compared to $\mathcal{L}_{b:c}$ of the single trilinear interaction given by Eq.~6
.

\subsection{Coherent and Incoherent Internal Driving}

In the main text, we focus on the incoherent resource of thermal states to generate coherent quantum phenomena. With coherent resources the amount of nonclassicality generated can be greatly increased, interestingly however the qualitative behaviour can already be observed for the thermal states. 
\begin{figure}[ht!]
    \centering
    \includegraphics{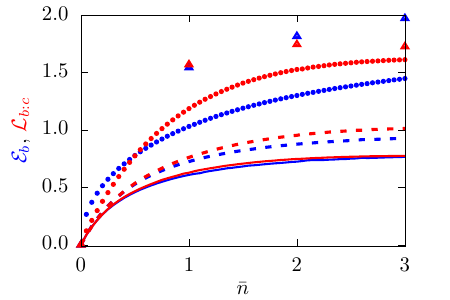}
    \vspace{-1ex}
    \caption{The role of different initial states on the entanglement production. $\mathcal{E}_b$ (blue) and $\mathcal{L}_{b:c}$ (red) are calculated using different resources to drive $a$. Represented are the thermal states (solid lines), coherent states (solid circles), Poissonian noise (dashed), and Fock states (solid triangles). $b$ is always initialised in the ground state. Note that for coherent states and Poissonian noise, we have $\bar{n} = \vert\alpha\vert^2$.}
    \label{fig:all_resources}
\end{figure}

In Fig.~\ref{fig:all_resources} we show the EP and LN with $a$ prepared in a coherent state (solid circles) $\ket{\alpha}$ with $|\alpha|^2=\bar{n}$. This is much greater than that achievable with thermal states (solid lines). If the phase is uncertain then the state has Poissonian noise in the Fock basis, and we obtain the phase randomised coherent states (PRCS) defined by
\begin{equation}
  \frac{1}{2\pi} \int_{0}^{2\pi} d\theta\, \ket{\alpha e^{i\theta}}\bra{\alpha e^{i\theta}} = e^{-\vert\alpha\vert^{2}} \sum_{n=0}^{\infty} \frac{\vert\alpha\vert^{2n}}{n!}\ket{n}\bra{n}.
\end{equation}
Here, it is instructive to compare with less noisy and more coherent drives under the same conditions. Poissonian noise (dashed lines) produces less nonclassicality and entanglement than the coherent state, but more than the thermal states. Furthermore, the Fock states, which have zero noise around the phonon number and no quantum coherence, but are extremely nonclassical, induce larger nonclassicality than the other examples provided. 

\begin{figure}[ht!]
   \centering
    \includegraphics{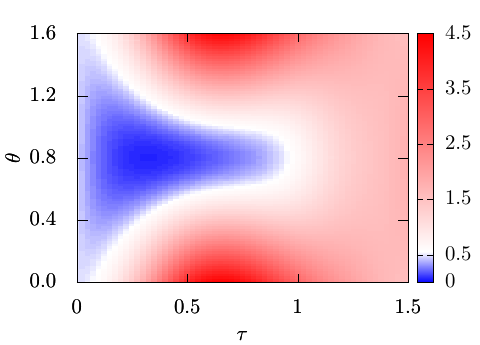}\\
    \vspace{-1.5ex}
    \caption{The variance of the generalised quadrature $X_\theta$ of mode $b$ as a function of scaled time $\tau = \Omega_T t$ and  phase $\theta$, when $a$ is driven coherently (with strength $\bar{n} = \vert\alpha\vert^2 = 3$).}
    \label{fig:sqz_rot_quad_cs}
\end{figure}

If mode $a$ is prepared in a coherent state then the nonclassical state prepared in $b$ is phase sensitive, in contrast with that generated by thermal energy, with the phase determined by the phase of the coherent state. For large $\alpha$ the mode $a$ in the trilinear Hamiltonian can be approximated with the c-number $\alpha=|\alpha|e^{i\phi}$ as 
\begin{equation}
    H\sim\alpha (b^\dagger)^2+\alpha^*b^2=|\alpha|\left(b^{\dagger\,2}e^{i\phi}+b^2e^{-i\phi}\right)\,.
\end{equation}
The coherent phase $\phi$ can be connected with the quadrature angle $\frac{\theta}{2}$ via $\phi=\theta-\frac\pi2$, so that the Hamiltonian takes the standard squeeze operator form
\begin{equation}
    H=-i|\alpha|\left(b^{\dagger\,2}e^{i\theta}-b^2e^{-i\theta}\right)\,.
\end{equation}

Fig.~\ref{fig:sqz_rot_quad_cs} shows the squeezing in the generalised quadrature $X_{\theta} = \frac{1}{\sqrt{2}} \left(b e^{-i\frac{\theta}{2}} + b^{\dagger}e^{i\frac{\theta}{2}}\right)$ for the coherent state with $|\alpha|^2=\bar{n}=3$ and $\phi=0$. The squeezing is most manifest along the quadrature axis defined by $\theta=\frac{\pi}{2}$, in accordance with the prediction of the saturated trilinear system. In contrast, for a phase random initial state such as a thermal state or PRCS, the nonclassicality is phase insensitive.


\subsection{\label{sec:decoherence} Influence of the Environment}

In our treatment of the trilinear dynamics, we have assumed unitary evolution due to the high quality of the dynamics and sufficient decoupling from the environment for both trapped ions and superconducting circuits over time required for such experiments. In this Section, we detail our numerical observations on the robustness of the thermally induced entanglement generation against such effects. If we ignore the contributions of the Lamb shift leading to a small renormalization of the system energy levels, the Lindblad master equation describing such environmental effects is given by 
\begin{align}
 \displaystyle\frac{d\rho_{S}}{dt}  = -i \left[H, \rho_{S}\right] + \sum_{k} \left( L_{k} \rho_{S} L_{k}^{\dagger} -  \frac{1}{2}\left[\rho_{S},\, L_{k}^{\dagger} L_{k}\right] \right).
 \end{align}
Here,  $\rho_{S}(t)$ is the reduced density matrix of the system. The operators $L_{k} = \sqrt{\lambda}_{k} A_{k}$ are the Lindblad jump operators, and the environment couples to the system through the operators $A_{k}$ with coupling rates $\lambda_{k}$. For the two mechanical modes coupled to the environment with same temperature $n_{\text{th}}$, we have the Lindblad operators ($L_k$'s) given by $\sqrt{\lambda_{a}(1+\bar{n}_{\text{th}})} \,a$, $\sqrt{\lambda_{a}\, \bar{n}_{\text{th}}}\, a^{\dagger}$, and $\sqrt{\lambda_{b}(1+\bar{n}_{\text{th}})} \,b$,  $\sqrt{\lambda_{b}\, \bar{n}_{\text{th}}}\, b^{\dagger}$. When considering the LN there are additional terms for the mode $c$ involved in the linear coupling. 

\begin{figure}[ht!]
  \centering
  \vspace{0ex}
    \includegraphics{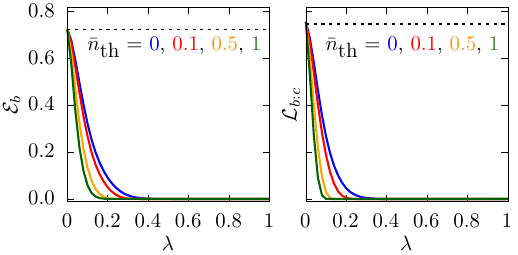}\\
    \vspace{-1.5ex}
    \caption{Effect of environmentally induced decoherence on $\mathcal{E}_b$ and $\mathcal{L}_{b:c}$ when all modes are connected to the same thermal bath with different temperatures $\bar{n}_{\text{th}} = 0$ (blue), 0.1 (red), 0.5 (yellow), and 1.0 (green). Initially, $a$ is in a thermal state with  $\bar{n}=2$, and $b$ and $c$ are in the ground states.  The black dashed lines represent the unitary evolution ($\lambda=0$). We fix the time as the optimised time of the unitary case, as described in the main text.}
    \label{fig:decoh_lambda}
\end{figure}

\begin{figure}[ht!]
  \centering
  \vspace{0ex}
    \includegraphics{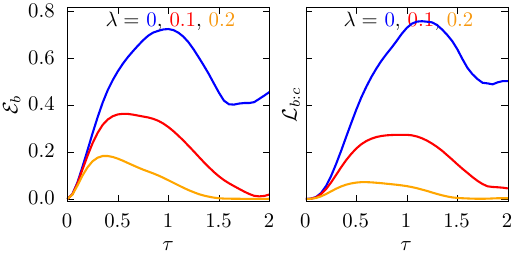}\\
    \vspace{-1.5ex}
    \caption{Effect of environmental induced decoherence on $\mathcal{E}_b$ and $\mathcal{L}_{b:c}$ when all modes are connected to the same bath with $\bar{n}_{\text{th}} = 0$. Initially, $a$ is in a thermal state with  $\bar{n}=2$, and $b$ and $c$ are in the ground state. We fix the time as the optimised time of the unitary case, as described in the main text.}
    \label{fig:decoh_time}
\end{figure}

In Fig.~\ref{fig:decoh_lambda} the effect of decoherence on both the EP and LN is examined in terms of the strength of the coupling to the environment. The system evolves with mode $a$ in a thermal state with $\bar{n}=2$ and the remaining modes in the ground state, until a fixed time set by the unitary evolution in the main text. Both quantities experience a sharp decay to zero with similar behaviour, with the rate increasing with the temperature of the environment. However in Fig.~\ref{fig:decoh_time} the direct evolution against time shows that while the nonclassicality is largely reduced, significant quantities are still visible, albeit at slightly shorter times. 

Importantly, as explained in the following section, the enhancement due to initial thermal noise remains visible despite the reduction in logarithmic negativity.

\subsection{Thermal Occupation of the Radial Mode}
\begin{figure}[ht!]
    \centering
    \includegraphics{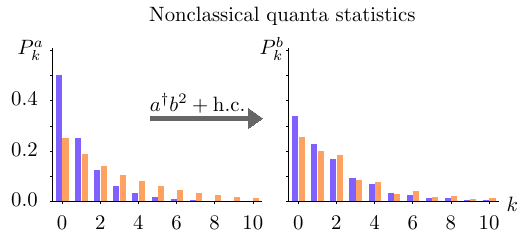}\\ \vspace{1ex}
    \includegraphics{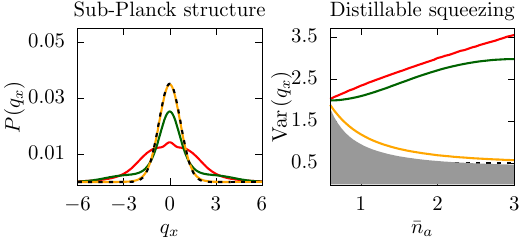}
    \vspace{-4ex}
    \caption{Analysis of the nonclassicality in mode $b$ as in Fig.~1 
    with $\bar{n}_b=1.5$. The clear visual generation of nonclassicality by the occupation of only even Fock states is lost, however, Klyshko's criteria still indicate nonclassicality. The distillable squeezing can also still achieve values below the shot noise by preparing $a$ in thermal states with $\bar{n}_a>3$.}
    \label{fig1_nb_bar_1_5}
\end{figure}

In our analysis of the nonclassicality and EP for the trilinear system, we assumed the idealisation that mode $b$ is prepared in the ground state with the only thermal energy in mode $a$. This is because both trapped ions experiments and superconducting circuits can achieve good ground states of these modes. However, to check the stability outside this limit we now briefly consider the effect of thermal occupation of modes $b$ and $c$. Advantageously, the thermally induced effects persist even if the thermal energy in mode $b$, characterised by an average occupation $\bar{n}_b$, is greater than 1. Fig.~\ref{fig1_nb_bar_1_5} shows a recreation of Fig.~1 
 from the main text with $\bar{n}_b=1.5$. The clear demonstration of nonclassicality from the phonon statistics in Fig.~1
is lost however Klyshko's criteria still verify nonclassicality. Similarly, the distillable squeezing in $q_x$ can still be reduced below the shot noise by very large thermal energy in $a$.

\begin{figure}[ht!]
    \centering
    \includegraphics{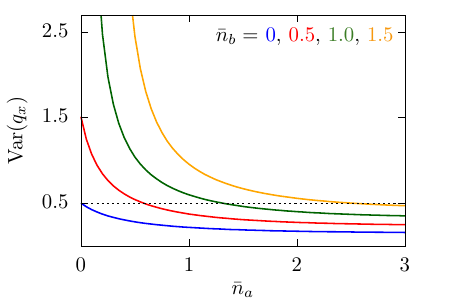}
     \vspace{-1ex}
     \caption{Universal distillable squeezing when mode $b$ deviates from the ideal ground state configuration to a thermal state with average phonon number $\bar{n}_b = 0$ (blue), 0.5 (red), 1 (green) and 1.5 (yellow) as a function of $\bar{n}_a$. The distillable squeezing indicates nonclassicality for large enough thermal energy in $a$.}
     \label{fig1:varx_nb}
 \end{figure}

In Fig.~\ref{fig1:varx_nb} we show how the distillable squeezing varies as a function of $\bar{n}_b$. The amount of thermal energy in $a$ required to see a reduction in the position variance below the shot noise increases with $\bar{n}_b$, however, it is still possible to see nonclassicality even for large values of $\bar{n}_b$.

\begin{figure}[ht!]
   \centering
    \includegraphics{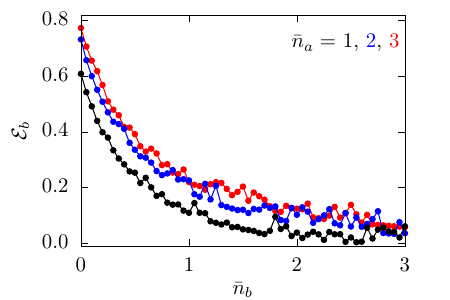}\\
    \vspace{-1.5ex}
    \caption{$\mathcal{E}_b$ as a function of the thermal occupation $\bar{n}_b$ of the radial mode $b$ for several values of $\bar{n}_a$. While the overall effect is to reduce the amount of EP produced, the EP is still visibly increased by an increase in the thermal occupation of mode $a$.}
    \label{fig:ep_n2bar}
\end{figure}

Finally in Fig.~\ref{fig:ep_n2bar} we also show the EP as a function of $\bar{n}_b$. Increasing $\bar{n}_b$ (Fig~\ref{fig:ep_n2bar}) drastically decreases the absolute value of the EP. However, our main result, that the EP is increased by increasing the thermal energy in mode $a$, is still visible up to saturation at $\bar{n}\approx3$.

\subsection{Effect of Free Motion on Nonclassicality and Entanglement}

Our analysis takes place in a frame of reference that eliminates the free motion of the component oscillators, which is suited to the platforms of trapped ions that we consider as well as similar systems such as superconducting circuits. If such trilinear Hamiltonians occur in a context in which such transformations cannot be performed then the effect of the free motion must be taken into account. The Hamiltonian, including the linear coupling required to study the LN, is modified to
\begin{equation}
    H_\text{FM}=\sum_{k=a,b,c}\omega_kk^\dagger k+\Omega_T(a^\dagger b^2+ab^\dagger{}^2)+g(b^\dagger c+bc^\dagger)\,.
\end{equation}

\begin{figure}[ht!]
   \centering
    \includegraphics{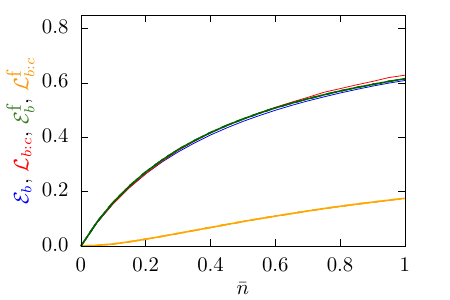}\\
    \vspace{-1.5ex}
    \caption{The entanglement potential and logarithmic negativity in the presence of free motion. Blue and red correspond to the EP and LN analysed in Fig.~2
    of the main text. The green and yellow correspond to the nonclassicality and entanglement dynamics involving the free motion. }
    \label{FreeMotion}
\end{figure}
We therefore briefly verify if the observed effects qualitatively persist if the free evolution is present. As in the main text, we select $\Omega_T=g$. Due to energy conservation, there is only one parameter for the frequency, as $\omega_a = 2\omega_b = 2\omega_c$. While the effects of free motion terms are typically destructive to quantum phenomena, here we find that the EP is in fact quite resilient to these effects but the LN suffers as shown in Fig~\ref{FreeMotion}. We note however that LN is still thermally stimulated and reaches observable values for experimental tests within this regime.

\subsection{Trilinear Hamiltonians with Trapped Ions}

A pair of ions with equal mass $m$ and charge $e$, contained in a harmonic trap, are coupled via the Coulomb interaction. If the radial trapping frequencies $\omega_x$ and $\omega_y$ are assumed to be much greater than the axial trapping frequency $\omega_z$, then the ions distribute themselves along the $z$-axis with equilibrium positions $(0, 0, z_{i}^{0})$~\cite{marquet_phononphonon_2003}, where $z_1^0 = -\sqrt[3]{\frac{e^2}{4m\omega_z^2}}=-z_2^0$. The Coulomb interaction can be expanded to second order, inducing a {\it natural} transformation to the normal modes of the motion, wherein the collective vibrational modes are decoupled from each other. The first step beyond this harmonic approximation, involving third order terms in the expansion, removes the decoupling of the spatial motion leading to interactions involving both the radial and axial modes. Components of these nonlinear interactions are resonant properties of the collective motion of the ions.

Recent experiments have achieved regimes in which these higher order interactions become relevant. The third order term in the Coulomb potential expansion is expressed, with $Q_\alpha =\frac12(r_{1\alpha}-r_{2\alpha})$ the relative position for axes $\alpha={x,y,z}$, as
\begin{equation}
    V^{(3)} = \frac{e^2}{4|Q_0|^4}Q_z \left(3\left(Q_x^2+Q_y^2\right) - 2Q_z^2\right)\,,
    \label{eqn:v3}
\end{equation}
where $Q_0=\frac12(z_1^0-z_2^0)$. Evidently, the relative motion of the ions becomes anharmonic and, importantly, the axial motion becomes coupled to the radial motion. From here, we will assume that the most significant components of the potential are those contributing to the interaction between the $x$ and $z$ spatial components. In this approximation, the potential energy can be expressed as
\begin{equation}
    V\approx\sum_{\alpha=x,z} m\Omega_\alpha^2Q_\alpha^2+\frac{3e^2}{4|Q_0|^4}Q_zQ_x^2\,,
\end{equation}
where we have defined the normal mode frequencies $\Omega_x^2=\omega_x^2-\frac{e^2}{4m|Q_z^0|^3}$ and $\Omega_z^2=\omega_z^2+\frac{e^2}{2m|Q_z^0|^3}$.

The motional variables can be quantized by promoting them to operators satisfying the canonical commutation relations $[\hat{Q}_\alpha, \hat{P}_\alpha] 
= i\hbar$, where $\hat{P}_\alpha$ 
is the momentum operator corresponding to the normal mode $\hat{Q}_\alpha$. 
The canonical variables of the relative motion can be linked to annihilation operators by the relations $\hat{Q}_z= \sqrt{\frac{\hbar}{2m\Omega_z}} (\hat{a} + \hat{a}^\dagger)$ and  $\hat{P}_z = i\sqrt{\frac{\hbar m\Omega_z}{2}} (\hat{a}^\dagger -  \hat{a})$, and $\hat{Q}_x= \sqrt{\frac{\hbar}{2m\Omega_x}} (\hat{b} + \hat{b}^\dagger)$ and  $\hat{P}_x = i\sqrt{\frac{\hbar m\Omega_x}{2}} (\hat{b}^\dagger -  \hat{b})$, with $[a,\, a^\dagger] = [b,\, b^\dagger]=\mathbb{I}$. Expanding $V$ in terms of the annihilation operators and moving to a frame rotating with the respect to the free motion generates an interaction Hamiltonian $H$ of the form
\begin{multline}
    H=\frac{3e^2}{4|Q_0|^4} \left(ae^{-i\Omega_zt}+a^\dagger e^{i\Omega_zt}\right)  \\ \times\left(b^2e^{-i2\Omega_xt}+b^\dagger{}^2e^{i2\Omega_xt} + b^\dagger b+1\right).
\end{multline}

When $\Omega_z = 2\Omega_x$, several terms become resonant and in the rotating wave approximation the interaction Hamiltonian takes the trilinear form
\begin{equation}
    H=\Omega_T\left(ab^\dagger{}^2+a^\dagger b^2\right),
\end{equation}
where $\Omega_T  =  \frac{3e^2}{4|Q_0|^4}$. For a more detailed Hamiltonian analysis see, for instance, Refs.~\cite{marquet_phononphonon_2003,lemmer_quantum_2018}. This is a partially degenerate trilinear interaction in which the creation (annihilation) of a phonon in the axial direction results in the annihilation (creation) of a pair of phonons in the radial direction. Alternatively for superconducting circuits trilinear interactions using SNAILs~\cite{frattini_3-wave_2017} can be considered to reach Eq.~(\ref{eqn:p_ham}).

\end{document}